\documentclass[sigconf]{acmart}
\usepackage{graphicx} 
\usepackage{hyperref}

\setcopyright{acmlicensed}
\copyrightyear{2026}
\acmYear{2026}
\acmConference{CHI '26 Workshop
on Tools for Thought}{April 16,
  2026}{Barcelona, Spain}
\date{April 2026}

\begin{document}

\title{seneca: A Personalized Conversational Planner}
\author{Simon Bohnen}
\email{simon.bohnen@tum.de}
\orcid{0000-0002-2549-0590}
\affiliation{%
  \institution{Technical University of Munich}
  \city{Munich}
  \country{Germany}
}

\author{Gabriel Garbers}
\email{gabriel.garbers@tum.de}
\affiliation{%
  \institution{Technical University of Munich}
  \city{Munich}
  \country{Germany}
}

\author{Lukas Ellinger}
\email{lukas.ellinger@tum.de}
\affiliation{%
  \institution{Technical University of Munich}
  \city{Munich}
  \country{Germany}
}

\author{Georg Groh}
\email{grohg@in.tum.de}
\affiliation{%
  \institution{Technical University of Munich}
  \city{Munich}
  \country{Germany}
}

\renewcommand{\shortauthors}{Bohnen et al.}

\begin{abstract}
Knowledge work demands sustained self-regulation, prioritization, and reflection—yet existing planning tools only partially support these needs. Digital to-do list applications feature task persistence but lack goal representation. Paper-based planning frameworks offer effective planning strategies but cannot adapt to individual users. Conversational AI systems enable flexible reflection but lack persistence and accountability. Moreover, none of these tools address a fundamental challenge: users' expressed demands often diverge from their underlying needs.

This paper introduces \textit{seneca}, a conceptual framework for a personalized, AI-assisted planner that integrates the complementary strengths of these three approaches. seneca combines a conversational agent that scaffolds reflection and asks clarifying questions, a persistent database that tracks goals and behavioral patterns, and a processor that synchronizes information between them. We describe this architecture and outline a phased evaluation strategy combining automated testing with simulated users and longitudinal human studies measuring goal attainment, planning realism, and goal–value alignment.
\end{abstract}



\received{12 February 2026}

\maketitle

\section{Introduction}

Knowledge work is becoming increasingly complex \cite{consoli_routinization_2023}. Thus, self-leadership — the ability to decide what to work on, align daily actions with long-term goals, and regulate attention in complex environments \cite{woods_learning_2023} — becomes more important. Existing productivity tools each address fragments of this challenge. However, none integrate reflection, prioritization, and reliable follow-through into a coherent system.

A further, often overlooked difficulty is that users' needs—such as goal clarity, value alignment, or reduced cognitive load—often differ from their expressed demands, such as adding a task or rescheduling a deadline \cite{pierce_undesigning_2014}. In current planning tools, tasks are captured as stated, without questioning whether they reflect the user's actual priorities or deeper intentions.

Motivated by these observations, we address two research questions:

\begin{quote}
\textit{What are the complementary strengths and shortcomings of existing planning tools?}
\end{quote}

\begin{quote}
\textit{How can we integrate the advantages of these existing tools into a single system?}
\end{quote}

This paper makes three contributions. In response to the first question, we identify the strengths and shortcomings of three existing approaches to planning support (\autoref{sec:existing-tools-and-their-shortcomings}). Addressing the second question, we propose a conceptual architecture—seneca—that combines these approaches into an integrated system designed to support clarification, reflection, and goal refinement (\autoref{sec:seneca}). Additionally, we outline an evaluation strategy for assessing whether such a system improves planning behavior (\autoref{sec:evaluation}).

\begin{figure*}[t]
  \centering
  \includegraphics[width=0.85\textwidth]{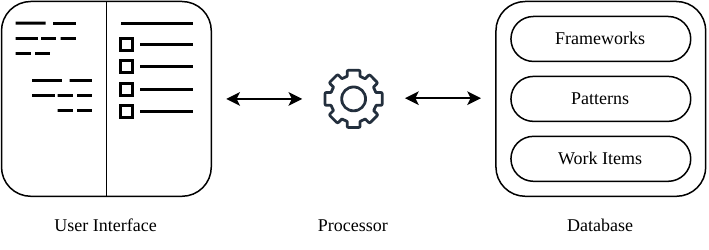}
  \caption{seneca's Core Components: The user interface combines a conversational agent with a structured work item view (left). The processor (center) synchronizes between the user-facing components and the database (right).}
  \label{fig:seneca-components}
\end{figure*}

\section{Existing Tools \& Their Shortcomings}\label{sec:existing-tools-and-their-shortcomings}

Productivity is often still understood through the lens of assembly-line work, where output depends on simple metrics such as time spent. Contemporary knowledge work, however, increasingly demands sustained self-regulation, prioritization, and reflection \cite{schkolski_influence_2025}. Being productive requires the ability to focus on what is important and urgent, while also making sound decisions about less immediately pressing tasks.

Existing planning and productivity tools only partially support these demands. In what follows, we review three influential approaches—digital to-do list applications, paper-based planning frameworks, and conversational AI systems—and identify the gaps that motivate seneca.

\subsection{Digital To-Do List Applications}

Classic digital to-do list applications such as Todoist, Apple Reminders, and Things address one core aspect of the problem space. They excel at reliable task capture, long-term persistence, and low-friction interaction. As external memory systems, they reduce cognitive load by allowing users to store and retrieve large quantities of tasks across time and contexts \cite{risko_cognitive_2016}. This reliability and scalability have made them foundational infrastructure for modern knowledge work.

However, these systems are largely content-agnostic and impersonal. Tasks are treated as isolated units unless manually organized, and the tools have no representation of user goals, values, habits, or working patterns. This lack of personalization can lead to users abandoning digital task management  \cite{haraty_how_2016}. Without adequate personalization, the gap between what the user enters and what the user actually needs cannot be addressed.

\subsection{Paper-Based Planning Frameworks}

Paper-based planning frameworks guide the user by incorporating specific planning philosophies into their structure. One such framework is Greg McKeown's Essentialism \cite{mckeown_essentialism_2014} and the associated 90-day planner. Essentialism emphasizes deliberate trade-offs, focus on a small number of meaningful objectives, and the rejection of non-essential commitments. Although this runs contrary to the assembly-line view of productivity, reducing goal quantity can decrease psychological distress \cite{gray_goal_2017}.

However, while these frameworks may be effective strategies overall, they cannot adapt to the user's specific needs. A paper planner cannot prompt reflection at crucial moments, nor can it provide encouragement or timely reminders. Moreover, paper-based planners offer no mechanism for detecting when a user's stated goals have drifted from their underlying values—a misalignment that may only become visible through active questioning or longitudinal tracking.

\subsection{Conversational AI Systems}

Conversational AI systems such as ChatGPT introduce a different interaction paradigm for cognitive support. Unlike both to-do list applications and paper-based planners, they enable natural-language interaction. Users can externalize thoughts, articulate uncertainty, and iteratively refine goals. From a cognitive perspective, these systems function as flexible thinking partners. They can scaffold reflection, reframe problems, and generate alternative perspectives. This aligns with theories of distributed cognition and cognitive offloading, in which reasoning is supported through interaction with external artifacts \cite{clark_extended_1998}.

However, conversational AI systems lack persistence and structure. Commitments are not reliably tracked over time. Accountability mechanisms are weak, as the AI does not actively follow up on prior conversations. Insights generated in one session are often disconnected from subsequent action \cite{liu_lost_2024, shi_large_2023}. Consequently, while conversational AI supports sensemaking and decision-making in the moment, it does not reliably translate cognition into sustained behavior.

Crucially, current conversational AI systems also do not systematically address the gap between what users say they want and what they actually need. Conversely, they often blindly obey the user's request or copy their biases \cite{greenblatt_alignment_2024, cherep_framework_2025}. Without mechanisms for clarification and goal refinement, conversational AI risks reinforcing surface-level demands rather than helping users uncover deeper intentions.

\section{seneca: A Conceptual Framework for AI-Assisted Planning}\label{sec:seneca}

The gaps identified in the previous section—lack of personalization in digital to-do lists, lack of interactivity in paper-based planners, and lack of persistence in conversational AI—motivate the design of an integrated system. seneca consists of three main parts: a database, a user interface combining a conversational agent with a structured work item view, and a processor that moves information between them (\autoref{fig:seneca-components}).

\subsection{The Database}

The database stores the user's work items, behavioral patterns, and pre-defined frameworks. Work items include tasks, goals, deadlines, completion statuses, and relationships between items. By storing these persistently, seneca retains the core advantages of conventional to-do list applications.

In addition, the database stores pre-defined planning and reflection frameworks. These include strategies such as time-boxing (e.g., the Pomodoro Technique), task limitation (e.g., essentialism), and prioritization heuristics (e.g., the Eisenhower matrix).

As the user interacts with seneca over time, these frameworks are augmented by user-specific patterns. These patterns personalize seneca, e.g. by changing its conversational style, modifying the frequency and duration of planning and reflection sessions, and adapting the triggers that cause seneca to nudge the user.

\subsection{The User Interface}

seneca's user interface has two complementary parts. The conversational agent allows the user to plan, execute, and reflect upon their work through text and voice—it is particularly suited for open-ended reflection and unstructured exchange. Alongside it, a structured work item view lets users directly browse, create, and edit tasks and goals, which is often faster and lower-friction for routine task management. The agent is designed to behave like a coach: it asks questions rather than providing fully formed plans. This ensures that the cognitive work remains with the user and promotes alignment between the user's implicit mental models and the externalized plans in seneca. It builds on metacognitive insights suggesting frequent overconfidence in what humans expect to accomplish within a certain time frame and what they actually achieve. seneca will hold up a mirror to contrast expectation and reality and visualize insights that help users increase planning realism skill. Beyond correcting biases, seneca's deliberate planning and reflection stages prompt the user's own critical thinking about their goals—for instance, recognizing that a task can be delegated or that two ostensibly separate goals share a common dependency.

This coaching stance is also the primary mechanism through which seneca addresses the need–demand gap. When a user adds a new task or expresses a goal, seneca can ask clarifying questions: \textit{Why is this important to you? How does this relate to your current priorities? Is this urgent, or does it merely feel urgent?} Over time, these interactions can help users develop a clearer understanding of their own intentions and motivate them to pursue their goals \cite{hudig_goalsetting_2025}.
Both components are visible simultaneously (\autoref{fig:seneca-components}): the conversational agent occupies roughly one third of the interface, while the remaining space shows structured views of sub-projects as expandable lists and tiles, including a permanent reminder of the highest-priority goal(s) for the current time period.

\subsection{The Processor}

seneca's processor moves relevant information between the conversational agent and the database. During a reflection session, the processor provides recent work items and the chosen framework's reflection template to the conversational agent. During and after the session, it stores new behavioral patterns and modified work items in the database. Similarly, the processor orchestrates how frameworks and patterns shape seneca's style and content during planning sessions.

The processor is tasked with bridging between the structured world of tasks and goals and the more fluid, text-based conversational agent. Keeping these two representations synchronized is essential for the system to function coherently. This synchronization challenge is non-trivial and represents a key technical design consideration for future implementation.

\section{Planned Evaluation}\label{sec:evaluation}

To assess whether seneca meaningfully supports planning and self-leadership, we propose a phased evaluation strategy.

\subsection{Phase 1: Automated Evaluation with Simulated Users}

In an initial phase, we plan to evaluate seneca's core mechanisms using simulated user interactions. These synthetic user profiles will be generated using specialized LLMs \cite{naous_flipping_2025}. This allows us to test whether seneca's clarifying questions lead to measurable changes in goal specificity, task structure, and plan coherence, without the confounds introduced by real user variability. Automated evaluation also enables rapid iteration on the system's prompting strategies and framework configurations.

\subsection{Phase 2: Human Evaluation}

In a subsequent phase, we plan to conduct a longitudinal study with real users. The evaluation will span six weeks and employ a two-pronged approach: (i) direct assessment of user activity (usage intensity and frequency over time) and (ii) indirect evaluation via user feedback, using both popup polls and $\leq$30 minute semi-structured interviews with 20 users. Key evaluation measures include:

\begin{itemize}
    \item \textbf{Goal attainment:} What proportion of goals set at the beginning of a planning cycle are ultimately achieved? This serves as a direct measure of whether the system helps users follow through on their intentions.
    \item \textbf{Planning realism:} Do users' estimates of task duration and workload capacity become more accurate over time? Improved realism would indicate that the system helps users develop a more honest understanding of their own work patterns, counteracting the planning fallacy \cite{buehler_exploring_1994}.
    \item \textbf{Goal–value alignment:} Do users report that their planned goals more closely reflect their underlying values and priorities after using the system? This subjective measure, captured through periodic self-report questionnaires, directly assesses whether seneca helps bridge the gap between expressed demands and actual needs.
\end{itemize}

These measures are chosen because they capture not just whether users complete tasks, but whether they become more self-aware, more realistic, and more intentional in their planning over time. Prior work on self-regulated learning and reflective practice has shown that such process-level measures are more informative than simple output metrics when evaluating cognitive support tools \cite{zimmerman_becoming_2002}.

\section{Discussion}

\subsection{Onboarding and Adaptation}\label{sec:onboarding-and-adaptation}

seneca asks users to engage in a form of epistemic labor—conscious planning and reflection—that may be unfamiliar to many. This process can feel slow, counterproductive, or even uncomfortable, as it makes both successes and perceived failures in planning more visible. A specific tension arises between seneca's intent to slow users down for reflection and the user's desire to quickly capture a task and move on; the structured work item view partially alleviates this by providing a low-friction path for routine interactions, while the conversational agent remains available for deeper engagement. Therefore, adapting seneca to the user's prior experience is necessary to support continued engagement. Research on habit formation and behavior change suggests that gradual exposure is more effective than immediate full-complexity immersion \cite{gibson_effects_2026}.

\subsection{Uncovering Behavioral Patterns and Refining Goals}

Through repeated reflection, users can uncover behavioral patterns in their own work. By becoming conscious of these patterns, users can learn to account for them in future planning. We hypothesize that this self-knowledge transfers: users who develop reflective planning skills with seneca may also plan more effectively with other tools or without any tool at all. Moreover, improved awareness of one's own habits and capacities can strengthen self-efficacy: users who understand their working patterns are better equipped to trust their own planning judgments. Notably, much of this self-knowledge is not created by seneca but was already latent; seneca merely makes it explicit. As a result, users may not attribute the improvement to the system—and may even believe they possessed this knowledge all along.

Frequent and honest reflection also improves the alignment between users' implicit goals and their explicitly stated goals. This is precisely the need–demand gap discussed in the introduction. Over time, seneca's clarifying questions and reflective prompts can help users articulate what they actually want, rather than what they initially say they want. This alignment prevents frustration and supports longer-term engagement. Additionally, a realistic and honestly constructed plan can reduce off-task rumination \cite{masicampo_consider_2011}.

\subsection{Social and Collaborative Considerations}

Similar to other AI-based tools for thought \cite{langer_why_2026}, using seneca carries the risk of reduced interaction with others. When planning and reflecting solely within seneca, users may become unaware of constraints and considerations of their peers. Several approaches can mitigate this risk. seneca could be used in team settings, for example during weekly planning meetings. It might also surface constraints and deliberations of other team members during a user's individual sessions, though this raises data privacy concerns. Lastly, seneca's coaching style can explicitly prompt users to consider the work and goals of others during planning and reflection.

\subsection{Evaluation Considerations}

The proposed evaluation strategy raises its own challenges. Simulated users in Phase 1 cannot capture the full complexity of real planning behavior, and predefined scripts may not reveal how users react to unexpected or unwanted clarifying questions. In Phase 2, goal attainment rates may be influenced by external factors such as changing work demands, making it difficult to attribute improvements solely to the system. Planning realism is similarly confounded, as users may become better estimators simply through experience rather than through seneca's interventions. Goal–value alignment, as a self-reported measure, is susceptible to social desirability bias and may not reflect actual changes in planning behavior. These limitations should be addressed through careful study design, including control conditions and qualitative interviews to complement quantitative measures.

\subsection{Limitations and Open Questions}

seneca is presented as a conceptual framework, not as an evaluated system. Several important questions remain open. How much clarification and questioning do users tolerate before the system feels intrusive? How should the system handle cases where the user's expressed preferences and inferred needs clearly conflict? And how can the system avoid imposing a particular normative framework on users who may not share its assumptions about what constitutes good planning? These questions require empirical investigation and motivate the evaluation strategy described above.

\section{Conclusion}

This paper introduced seneca, a conceptual framework for a personalized, AI-assisted planner that synthesizes conversational AI, established planning frameworks, and persistent task management. seneca is designed to help users bridge the gap between their expressed demands and their underlying needs through clarifying questions, structured reflection, and gradual goal refinement.

Several open challenges remain. Users may find active questioning intrusive, particularly when the system surfaces conflicts between expressed preferences and inferred needs. Over-reliance on an individual planning tool risks reducing collaboration and awareness of others' constraints. Onboarding requires careful scaffolding to avoid discouraging users unfamiliar with structured reflection. Addressing these challenges empirically is the focus of our planned evaluation, which combines automated testing with simulated users and longitudinal studies measuring changes in reflection frequency, task completion, and planning behavior.

\bibliographystyle{ACM-Reference-Format}
\bibliography{references}

\end{document}